\documentclass{emulateapj}
\usepackage{multirow}
\usepackage{tabularx}

\shorttitle{Distribution of Coalescing Compact Binaries}
\shortauthors{L.Z. Kelley et al.}

\begin{document}

\title{The Distribution of Coalescing Compact Binaries in the Local Universe: Prospects for Gravitational-Wave Observations}

\author{Luke Zoltan Kelley\altaffilmark{1}, Enrico Ramirez-Ruiz\altaffilmark{1}, Marcel Zemp\altaffilmark{2}, J\"{u}rg Diemand\altaffilmark{3}, and Ilya Mandel\altaffilmark{4}}
\altaffiltext{1}{Department of Astronomy and Astrophysics, University
  of California, Santa Cruz, CA 95064}
\altaffiltext{2}{Department of Astronomy, University of Michigan, Ann Arbor, MI 48109}
\altaffiltext{3}{Institute for Theoretical Physics, University of Zurich, 8057 Zurich, Switzerland}
\altaffiltext{4}{NSF Astronomy and Astrophysics Postdoctoral Fellow, MIT Kavli Institute, Cambridge, MA 02139}

\begin{abstract}
Merging compact binaries are the most viable and best studied candidates for gravitational wave (GW) detection by the fully operational network of ground-based observatories.  In anticipation of the first detections, the expected distribution of GW sources in the local universe is of considerable interest. Here we investigate the full phase space distribution of coalescing compact binaries at $z = 0$ using dark matter simulations of structure formation. The fact that these binary systems acquire large barycentric velocities at birth (``kicks") results in merger site distributions that are more diffusely distributed with respect to their putative hosts, with mergers occurring out to distances of a few Mpc from the host halo. Redshift estimates based solely on the nearest galaxy in projection can, as a result, be inaccurate. On the other hand, large offsets from the host galaxy could aid the detection of faint optical counterparts and should be considered  when designing  strategies for follow-up observations. The degree of isotropy in the projected sky distributions of GW sources is found to be augmented with increasing kick velocity and to be severely enhanced if progenitor systems possess large kicks as inferred from the known population of pulsars and double compact binaries. Even in the absence of observed electromagnetic counterparts, the differences in sky distributions of binaries produced by disparate kick-velocity models could be discerned by GW observatories,  within the expected accuracies and detection rates of advanced LIGO--in particular with the addition of more  interferometers.
\end{abstract}

\email{lzkelley@ucsc.edu}

\keywords{gravitational waves \textemdash \textrm{ }stars: neutron \textemdash \textrm{ }binaries: general}


\section{Introduction}
The merger of double compact objects represents the first identified and most predictable source of gravitational wave (GW) radiation \citep[e.g.][]{phi91}.  Only recently have the first GW observatories come online, and the first detection events are expected in the next few years.  Over the past three decades the merger rate within the local universe has been thoroughly examined \citep[see e.g.][for recent reviews]{lig10, man10b}.

The merger rates are expected to be dominated by mergers of neutron-star binaries, with $\langle{\Re}\rangle\sim 1\,\textrm{Mpc}^{-3}\,\textrm{Myr}^{-1}$.  However, these rates are significantly uncertain, since they come either from extrapolations from the small observed sample of Galactic binary pulsars whose luminosity distribution is not well constrained, or from population-synthesis models that have many ill-determined parameters such as common-envelope efficiencies.  In particular, \citet{lig10} estimate the confidence bounds on the neutron-star binary merger rates as $\Re \approx 0.01-10\, \textrm{Mpc}^{-3}\,\textrm{Myr}^{-1}$.  The horizon distances\footnote{The horizon distance is the maximum distance at which a signal can be detected with a given signal-to-noise threshold (e.g., 8); for a single detector, this is the distance at which gravitational waves from a face-on, overhead binary can be detected.} for the initial and advanced LIGO/Virgo detector networks are estimated as $\mathcal{D} \sim 30$ and $\sim$ 400 Mpc, respectively, based on the distance at which a single detector could detect gravitational waves from a neutron-star binary at a signal-to-noise ratio of 8.  \citet{lig10} estimate that the advanced LIGO/Virgo network could plausibly detect between 0.4 and 400 neutron-star binaries per year, with a likely rate of approximately 40 detections per year.

The prospects for detection and characterization of GW sources are thus sensitive to the distribution of compact binaries in the local universe (i.e. at distances $\leq \mathcal{D}$). The fact that these systems must have large systemic velocities at birth \citep{bra95, fry97} implies that by the time they merge, after approximately a Hubble time, they will be far from their birth sites. The locations of merging sites depends critically on the binary's natal kick velocity and the temporal evolution of the gravitational potential of the host halo as well as that of its nearby neighbors \citep{zem09}.

In this {\it Letter}, we study the evolving distribution of compact binary systems from formation until coalescence at $z= 0$ using cosmological simulations of structure formation.  This allows us to examine the full radial and angular distributions of merging compact binaries in the local universe. The organization is as follows.  In \textsection \ref{sec_methods}, we describe the numerical methods and initial setup and the criteria used to select a local-like universe. The distributions of compact binaries at $z= 0$ are presented in \textsection \ref{sec_distribution} for three different kick velocity scenarios, and in \textsection \ref{sec_distinguish} we examine the ability of GW observatories to discern between them experimentally.  Finally, \textsection \ref{sec_discussion} discusses the implications of our findings.

\newpage
\section{Methods and Initial Model}
\label{sec_methods}

\subsection{Simulation}
\label{sec_sim}
The focus of this work is to understand the distribution of compact binaries in the local universe using cosmological simulations. To this end, we have performed a dark matter only cosmological structure formation simulation following the numerical procedure outlined in \citet{zem09}.  A comoving 80 Mpc periodic box is initialized at redshift $z = 22.4$ (161 Myr) and uses WMAP3 cosmological parameters \citep{spe07}.   The initial conditions are evolved using the parallel tree code PKDGRAV2 \citep{sta01} until $z = 1.60$ (4.24 Gyr).  At this time, we populate each halo with mass greater than $2.15\times10^{11} M_\odot$ (of which there are 2461 in the simulation) with 2000 massless tracers. 

Each tracer particle is meant to represent a compact binary system, which, on average, forms around the peak of the star formation epoch \citep{mad96,mad98}.   In general, the local merger rate is given by the convolution of the star formation rate with the probability distribution $P(\tau)$ of the merging time delays $\tau$.  Compact binaries formed at the peak of the star formation history, merging after delays consistent with the orbital separations of known relativistic binary pulsars \citep{osh08}, dominate the local merger rate\footnote{For $P(\tau)\propto 1/\tau$ this early-assembled population could increase the local event rate by at least $\sim 3$ \citep{gue05}.}.

Tracers are injected into the center of their halo, with an isotropic Maxwell-Boltzmann velocity distribution with mean speeds $\bar{v} = 360$, 180, and 90$\textrm{ km s}^{-1}$ and dispersions $\sigma = 150$, 75, and 37.5 $\textrm{ km s}^{-1}$ (hereafter denoted as models $M_{360}$,  $M_{180}$ and  $M_{90}$). This is consistent with the magnitude of the natal kicks required to explain the observed parameters of binary NS systems \citep{bra95,fry97} --- only when the center of mass kicks have magnitudes exceeding 200 km s$^{-1}$ can the progenitor orbits be sufficiently wide to accommodate evolved helium stars and still produce the small separations measured in these systems.

The contribution of individual tracers to the overall population is weighted linearly with their progenitor halo's mass (at $z=1.6$) in all of our calculations. 

Finally, the cosmological box together with the tracer particle populations are evolved until redshift $z= 0$ (13.8 Gyr). This results in diverse predictions of compact binary demographics at $z=0$ in the case of an isotropic  kick velocity distribution whose properties do not vary with the initial binary separation. Merger times in population synthesis models are found to be relatively insensitive to the initial kick velocity \citep[e.g.,][]{blo99}. This not only justifies our assumption but, when  taken together with the progenitors' long time delays \citep{osh10}, also reinforces the validity of a single injection time.

\subsection{Local-Like Universe Selection}
\label{sec_sel}
Once the tracers and DM are evolved to $z=0$, a local-like universe is selected based on criteria adapted from \citet{hof08}.  The local-like universe is characterized here by the following:
\renewcommand{\labelenumi}{\roman{enumi}.}
\begin{enumerate}
\item
There are two dark matter halos, representing the Milky-Way and Andromeda pair, with maximum circular velocities $V_c \in [125, 270]$ $\textrm{ km s}^{-1}$.
\item
These halos are separated by $d \leq 1.4$ $h^{-1}\textrm{Mpc}$, and approaching each-other (i.e. $\dot{d} \leq 0.0 \textrm{ km s}^{-1}$).
\item
There is a Virgo-like halo at a distance $d \in [5, 12]$ $h^{-1}$Mpc, and $V_c \in [500, 1500]$ $\textrm{ km s}^{-1}$
\item
No halos with comparable or higher maximum circular velocities than either of the pair exist within $3$ $h^{-1}$Mpc, and no other Virgo-like halos exist within $12$ $h^{-1}$Mpc.
\end{enumerate}

The first three constraints resulted in three local-like groups.  Inclusion of the fourth criterion resulted in a single, optimal environment for our analysis. These criteria ensure that the evolutionary environment is similar to that of the actual Milky Way and local group galaxies.

\section{The Local Distribution of Compact Binaries}
\label{sec_distribution}
\begin{figure}[h]
\begin{center}
\vspace{1in}
\includegraphics[width=\textwidth]{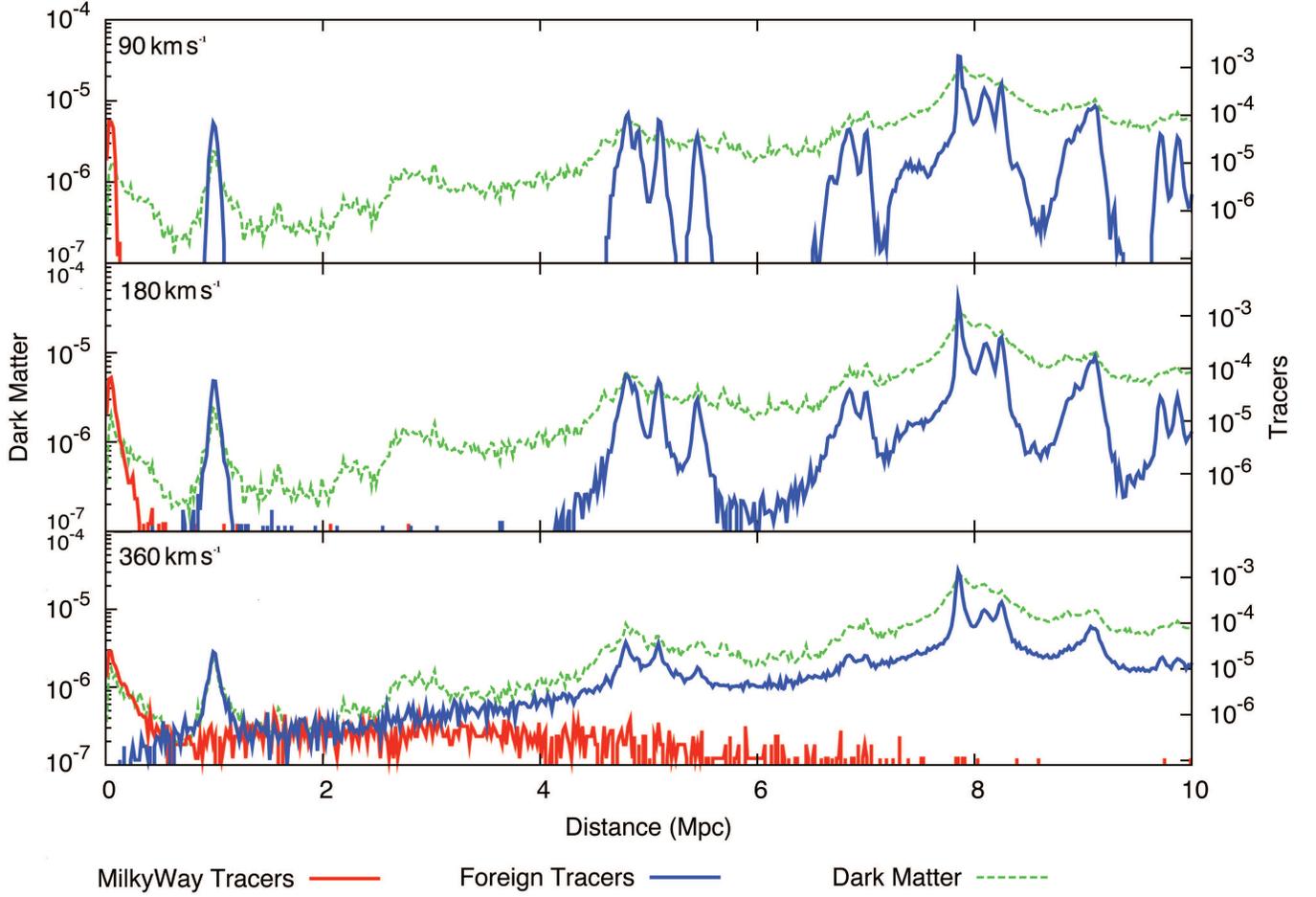}
\end{center}
\caption{Tracer vs. dark matter distribution in a local-like universe as a function of barycentric kick-velocity.  Integrated particle mass in uniform radial-width shells is plotted versus distance from a solar-equivalent offset from the Milky Way center.  The vertical axes are plotted in arbitrary units of number per unit length, with tracers normalized with respect to the total tracer population as described in \S \ref{sec_sim}.  As the kick-velocity increases from 90 $\textrm{km s}^{-1}$ (top panel) to 360 $\textrm{km s}^{-1}$ (bottom panel), a noticeable portion of tracers becomes delocalized, forming a background and mixing populations.}
\label{fig_radial}
\end{figure}

We now examine the local, three-dimensional distribution of merging compact binaries, which are characterized here by the massless tracer particle population at $z=0$ centered on the Milky Way-like galaxy as defined in \S \ref{sec_sel}.  Figure \ref{fig_radial} shows the radial distribution of tracers and dark matter within our local-like universe.  In models $M_{180}$ and $M_{90}$, tracer particles closely follow the dark matter central-density peaks, just like the galaxies themselves in CDM cosmology \citep{blu84}.

As the natal, barycentric kick-velocity becomes comparable to the escape velocity of the progenitor halos, an increasing fraction of tracers escape.  These unbound tracers pollute the intergalactic region, thus forming a particle background which closely follows the overall dark matter distribution, as seen for model $M_{360}$ in Figure \ref{fig_radial}.  In super-galactic regions with more continuous gravitational potentials (i.e. more densely and uniformly populated halo environments, such as clusters), the tracer background becomes more heavily populated.  The relative isolation of the Milky Way-Andromeda pair contributes to the enduring presence of strong native tracer peaks located at each halo center.  As a result, the extended background distributions of tracers--centered on the pair--are only apparent in the highest kick velocity model.

\begin{figure}[h]
\begin{center}
\vspace{1in}
\includegraphics[width=\textwidth]{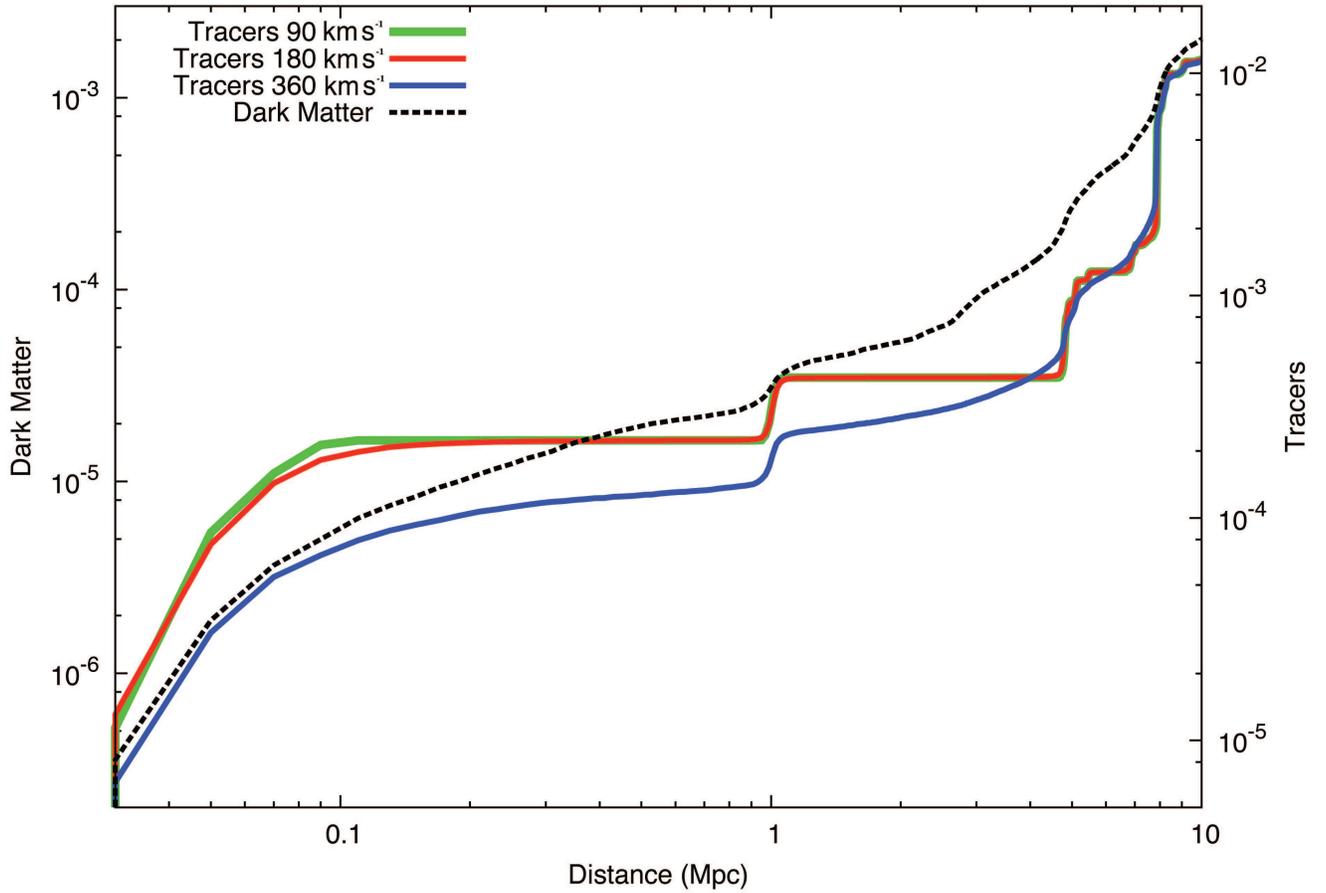}
\end{center}
\caption{Cumulative distribution of tracers and dark matter within a given distance as a function of kick-velocity.  The vertical axes are plotted in arbitrary units of number per unit volume, with tracers normalized self-consistently.  Although the number of tracers in the central halo is noticeably lower for the highest kick-velocity model, the difference is negligible once the volume reaches the Virgo-like cluster, where the background distribution of tracers outweighs changes in local distributions.  The number of merging binaries is assumed to be proportional to the mass of the host halo. }
\label{fig_rates}
\end{figure}

As seen in Figure \ref{fig_rates}, the number of tracers within a sphere encompassing the Milky Way and Andromeda halos is noticeably depleted at higher kick-velocities.  At these velocities, the host halos are unable to effectively retain most of their tracers.  As the sphere's volume approaches the Virgo cluster, the number of (sub)halos becomes so large that the mean separation between central peak densities decreases below the characteristic size of the background tracer population.  As a result, the variation with kick-velocity in the tracer distributions is drowned-out. It should be noted that the expected event rate in such a small volume is negligible \citep{lig10}, and thus the effects of varying kick-velocity will be indiscernible in the integrated merger-rate of compact objects within LIGO/Virgo detection horizons.

\begin{figure}
\begin{center}
\vspace{1in}
\includegraphics[width=\textwidth]{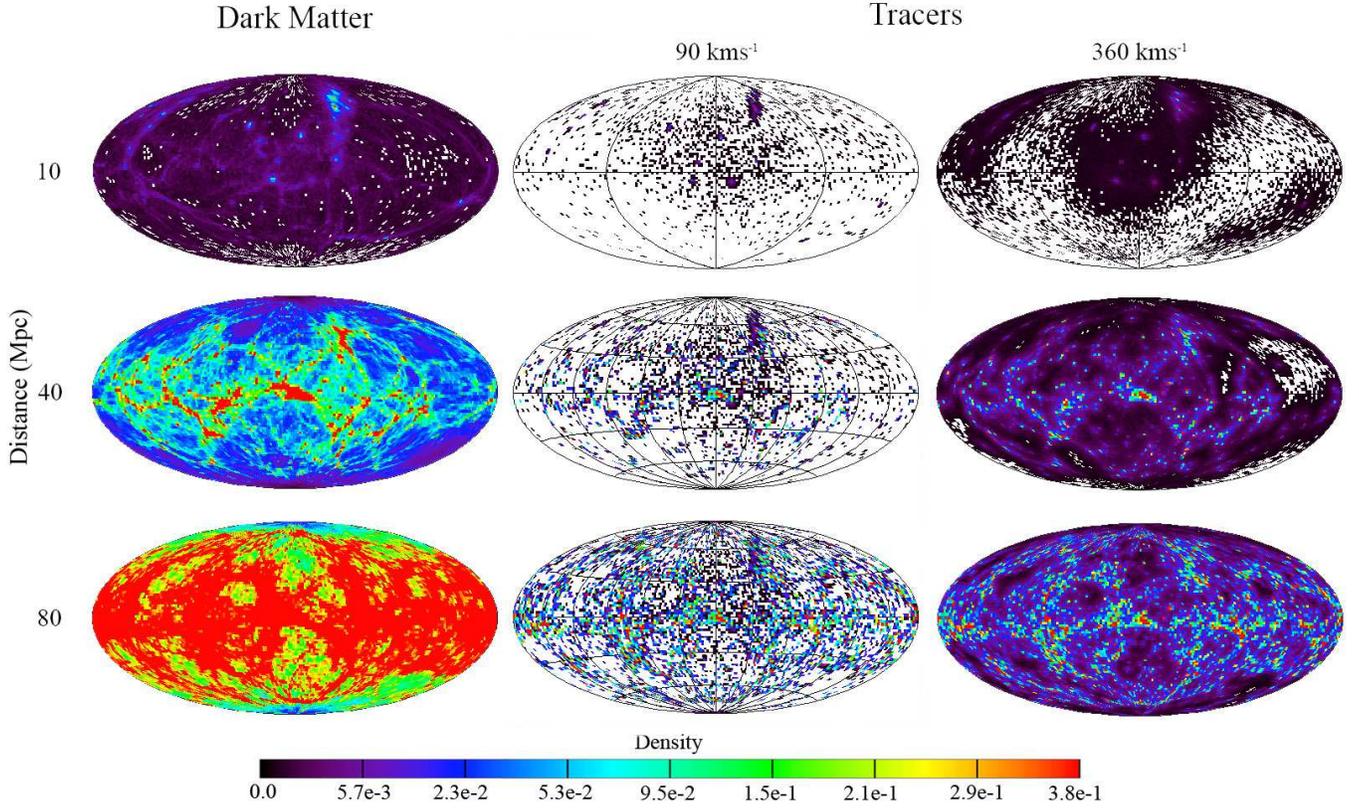}
\end{center}
\caption{Sky maps of dark matter (first column) and tracers with highest and lowest kick velocity scenarios (second and third columns, respectively) as a function of distance.  Figures make use of Hammer projections with $2^{\circ}$x $2^{\circ}$ bins.  Densities are plotted in units of column density, scaled to the maximum densities of the dark matter and normalized tracer distributions independently.  Pixels with no tracers or dark matter are white, corresponding to densities less than the resolution of the simulation.  Although tracer peak densities remain relatively unchanged, a tracer-background forms as the kick velocity approaches the escape velocity.  At 90 $\textrm{km s}^{-1}$ the tracers follow only the dark matter overdensities, as does the light-distrubtion.  The distributions approach isotropy slower than the dark matter distribution.  Differences in distribution are clearly apparent.  Note the logarithmic color scale.}
\label{fig_map}
\end{figure}

Although the integrated tracer distribution is insensitive to the model, the angular distribution of tracers depends strongly on the binary's kick velocity.  This is evident in Figure \ref{fig_map}, which plots sky maps of tracers and dark matter within a given volume (resolved to 4 square degree pixels). As expected, high velocity kicks lead to more pronounced isotropies when compared to the low velocity kick scenarios.  At 10 Mpc, $\sim$40\% of the $M_{360}$ weighted tracers lie in pixels outside those of $M_{90}$; this fraction falls to 15\% and 10\% for 40 and 80 Mpc respectively.  This trend results from the increasing isotropy of dark matter in projection at progressively larger scales.  

For large velocities, the distribution of GW sources forms a sky continuum (Figure \ref{fig_map}) rather than well-isolated substructures --- complicating  host galaxy identification and thus redshift determination.  On the other hand, the extension of the particle tracer distributions, which grows with increasing kick velocity, could aid the detection of photonic counterparts, especially at optical wavelengths. This is because at large kick velocities the majority of the mergers will take place well outside the host galaxy's half-light radius.

\section{Predictions for Gravitation Wave Observations}
\label{sec_distinguish}
The number of detections required for GW observatories to be able to reconstruct the kick-velocity distribution is examined here. 
Timing triangulation from relative GW phase shifts\footnote{For a review of GW emission from compact binaries see \cite{hug09}.} between widely separated detectors is the primary source of sky localization \citep{fai09}, and Fisher matrix or Markov Chain Monte Carlo techniques can be used to  compute error estimates \citep{slu08}.  The intricacies of parameter determination and error estimation can be extensive, as correlations between waveform parameters mean that some parameters (such as distance and inclination) are partially degenerate \citep[see, e.g.][and references therein]{cut94}. Typically, with a three-detector network, the distance to the GW event could be determined to within a $\sim$20--50\% uncertainty, and the angular location of the event to $\sim 5-50$ square degrees, depending on source location, masses, and signal-to-noise ratio \citep{fai09, slu08}.  The large uncertainty in the distance determination can be understood when considering its dependence on the signal amplitude, which is much more uncertain than the phase-space information.

To estimate the number of events required to distinguish between different kick velocity models, we apply a Bayesian approach similar to that used by \citet{man10} to approximate the efficacy of population reconstruction from GW signals.  Data sets $D_j \in D_{360}, D_{180} \;\textrm{and}\; D_{90}$ are drawn from each model $M_i \in  M_{360}, M_{180}  \;\textrm{and}\;  M_{90}$ respectively.   Each data set contains $n$ independently drawn data points (i.e. tracers), characterized by 3 position coordinates; i.e. $D_i (n) = [x_{i,1}(r,\alpha, \delta), x_{i,2}(r,\alpha, \delta), \ldots x_{i,n}(r,\alpha, \delta) ]$.  The probability of a tracer being selected for a given data set is linearly proportional to the halo mass of the progenitor, in accordance with the weighting scheme of \S \ref{sec_sim}. The probability that a particular model $i$ fits a data set $j$ can be rewritten using Bayes' formula: 
\begin{equation}
P( M_i | D_j(n) ) = \frac{ P(D_j(n) | M_i) \cdot P(M_i) }{ P(D_j(n))}.
\label{eq_bay1}
\end{equation}

Throughout our analysis we assume flat priors [$P(M_i) = P(M_j)$], and equivalent evidence [$P(D_i) = P(D_j)$].  A comparison between models then yields:
\begin{equation}
\frac{ P( M_i | D_i(n) ) }{ P( M_j | D_i(n) ) } = \frac{ P(D_i(n) | M_i)  }{ P(D_i(n) | M_j)  } = \displaystyle\prod_{k=1}^n \frac{ P( x_{i,k} | M_i)  }{ P( x_{i,k} | M_j)  },
\label{eq_bay2}
\end{equation}
where the probability of a particular data point given a specific model, $P( x_{i,k} | M_j)$,  is described by the convolution of the point spread function (PSF--$S$) of the detector with the probability distribution function of the model in question. That is 
\begin{equation}
P( x_{i,k} | M_j) = \displaystyle\sum_{l=1}^q S(x_{i,k} | \textrm{\small{pixel}}_l ) \cdot P(\textrm{\small{pixel}}_l | M_j),
\label{eq_bay3}
\end{equation}
where the sum is being performed on each pixel ($\textrm{\small{pixel}}_{l}$) for all $q$ pixels.

The PSFs are assumed here to be gaussian in each coordinate direction, characterized by standard deviations in distance, right ascension, and declination: $\sigma_{\rm high} = [5\%, 1^\circ, 1^\circ]$,  $\sigma_{\rm med} = [30\%, 2^\circ, 2^\circ]$, $\sigma_{\rm low} = [50\%, 4^\circ, 4^\circ]$.  These reflect different assumptions for the high, medium, and low accuracy of positional reconstruction for gravitational-wave detections.  The exact parameter-estimation accuracy is difficult to predict, since it will depend both on the details of the detector network (e.g., the relative sensitivity of detectors and their calibration accuracy) and on the specifics of individual events (their signal-to-noise ratio, and the sky location and orientation of the binary).  Therefore, these three assumptions should be considered only as possible predictions for typical accuracies.  Thus, low accuracies may be typical for events detected with a three-detector LIGO/Virgo network at the threshold of detectability.  Meanwhile, the addition of a fourth interferometer, such as a possible AIGO detector in Australia or LGCT in Japan, could significantly enhance the sky localization accuracy and moderately improve distance sensitivity (Fairhurst et al. 2010)\nocite{fai10}, making medium-accuracy measurements typical and high-accuracy measurements possible. 

In these calculations, we compare hypothetical GW observations with models of compact binary distributions.  This comparison is being made assuming that the local dark matter distribution is perfectly known.  In reality, this is not the case; and the results presented here are thus optimistic.  In the future, the comparison between model and observation should be refined to include the local distribution of light (e.g. galaxies) rather than dark matter halos.

Table \ref{tab_bay} summarizes the ability of GW observatories to discern the kick velocity distribution of the merging binaries from the reconstructed angular positions and distances (assumed to be determined without a galaxy host association).  Two sample volumes are considered: 40 and 80 Mpc.  This is done in order to understand the sensitivity of our results to the uncertainty in physical separation which, for a fixed angular resolution, varies with distance.  We find that $\sim50$ events are required to distinguish between the lowest and highest kick velocity scenarios for moderate detector accuracies, irrespective of which sample volume is examined.  For low detector accuracies, $50-350$ detections are necessary\footnote{
It is important to note that the number of detections required is highly sensitive to the model from which the data is drawn, not simply on which models are being contrasted.}.  Thus, a distinction between the two extreme models is possible once advanced detectors come online, with an expected event rate of $\sim$ 40 per year for detections of binary neutron star mergers \citep{lig10}.\footnote{The event rate estimates have significant uncertainties, and range from pessimistic estimates of $\sim$0.4 events per year to optimistic estimates of $\sim$400 events per year \citep{lig10}.}   Meanwhile, distinguishing between the two low-kick scenarios is very difficult, if not impossible, until the era of third-generation detectors. The addition of a fourth GW detector to the LIGO/Virgo network would significantly improve source localization, and thereby the accuracy with which event distributions could be distinguished.

Assuming a LIGO/Virgo horizon of $\sim 400$ Mpc, only $\sim$ 10\% of all detected mergers would take place within 80 Mpc.  With a constant angular resolution, the uncertainty in physical position is proportional to the event's distance, suggesting that using events at greater distances leads to a degradation in the ability to distinguish between kick-velocity models.  Although we find no clear increase in the number of required detections between the 40 and 80 Mpc samples, further investigation is required to assess the effects of a larger sample volume.


\section{Summary}
\label{sec_discussion}
In this \textit{Letter}, we use dark matter cosmological simulations to examine the full three-dimensional distribution of coalescing compact binaries in the local universe under the following assumptions. First, we assume a single epoch of star formation and a simple star formation recipe; that is, the contribution of a particular halo to the total star formation is directly proportional to its dark matter mass.  Although a more realistic treatment of star formation should be considered, we do not expect that our qualitative results will change significantly.  Second, we assume an isotropic natal kick velocity distribution, whose properties are invariant of initial binary separation.  Under this assumption, the merging time is independent of the kick velocity.  This is found to be a reasonable approximation in binary population synthesis models, which helps justify our single epoch of tracer injection. 
Third, our comparisons between kick velocity models in \S \ref{sec_distinguish} assume a perfect knowledge of the local dark matter distribution, when in actuality this distribution would have to be deduced from the observable, local universe.  Finally, due to computational constraints, only an 80 Mpc region of the expected 400 Mpc horizon of advanced LIGO/Virgo has been modeled.  Despite the increased uncertainty in the true-distance offset between host and merger at such distances, the difference between our 40 and 80 Mpc results (Table \ref{tab_bay}) suggests that our methods could remain effective in deducing the kick velocity distribution with a reasonable number of detections.  Keeping these assumptions in mind, it is still evident that the use of static, non-evolving potentials for individual hosts at the time of binary formation severely overestimates the retention of all but the lowest barycentric velocity systems \citep{fry99, bel00, ros03, blo99, bul99, por98}.  

Static calculations predict that the distribution of gravitational wave sources in the sky should closely trace the distribution of galaxies.  An accurate inclusion of evolving host halo potentials in cosmological simulations have shown this to be inaccurate \citep{zem09}.  In fact, we show that not only do the distributions of merging compact binaries extend well beyond their birth halo, but variations in kick velocity lead to marked differences in their sky distributions.  The repercussions of this result are twofold.  On one hand, we find that the variation in the projected distribution of double compact objects with different natal kick-velocities should be distinguishable with the expected accuracies of GW observatories.  In principle, this will allow important information on the formation and evolution of the binary progenitor to be deciphered from the distribution of GW detections alone.  On the other hand, the fact that the distribution of merging binaries does not accurately trace the locations of their birth halos complicates redshift determination.  Having said this, the presence of a binary distribution extending well beyond the half-light radius of their hosts suggests that associating optical counterparts to GW events could be easier as they are less likely to be drowned out by their host galaxy's light.  This is particularly important as the optical counterparts are predicted to be relatively dim \citep{li98, ros02, kul05, met10}.

Gravitational waves offer the possibility of casting proverbial light on otherwise invisible phenomena;  they will -- by their very nature -- tell us about events where large quantitites of mass move in such small regions that they are utterly opaque and forever hidden from direct electromagnetic probing \citep[see, e.g.][]{lee07}.  A time-integrated luminosity of the order of a fraction of a solar rest mass is predicted from merging compact binaries.   Ground-based facilities, like LIGO, GEO600 and Virgo, will be searching for these stellar-remnant mergers in the local universe.   The distribution of merger sites is thus of considerable importance to GW observatories.   The proposed use of galaxy catalogs as priors when passing triggers from possible GW detections to point telescopes for electromagnetic follow-ups will need to account for the possibility of mergers away from the observed galaxies.  Using cosmological simulations of structure formation, the local sky distributions are found to vary with the kick velocity distributions of the progenitor systems, allowing a determination of the cosmography of massive binary stars.  Despite the fact that individual detections lack the positional accuracy of electromagnetic observations, it may be possible to strengthen the case for (or against) high natal kick velocities based solely on GW observations.  The addition of more gravitational-wave detectors to the LIGO/Virgo network will greatly improve our ability to distinguish between models with different kick velocity distributions by improving the positional reconstruction of individual events.


\newpage
\acknowledgments
We thank C. Fryer, V. Kalogera and R. OÕShaughnessy for useful discussions and the referee for constructive comments.  We acknowledge support from NASA NNX08AN88G and NNX10AI20G (L.Z.K. and E.R.), the David and Lucile Packard Foundation (E.R.); NSF grants: AST-0847563 (L.Z.K. and E.R.), AST-0708087 (M.Z.), AST-0901985 (I.M.); and the Swiss National Science Foundation (J.D.).   Computations were performed on the Pleaides UCSC computer cluster.


\bibliographystyle{apj}
\bibliography{biblio}

\begin{thebibliography}{32}
\expandafter\ifx\csname natexlab\endcsname\relax\def\natexlab#1{#1}\fi

\bibitem[{{Abadie} {et~al.}(2010){Abadie}, {Abbott}, {Abbott}, {Abernathy},
  {Accadia}, {Acernese}, {Adams}, {Adhikari}, {Ajith}, {Allen}, {Allen},
  {Amador Ceron}, {Amin}, {Anderson}, {Anderson}, {Antonucci}, {Aoudia},
  {Arain}, {Araya}, {Aronsson}, {Arun}, {Aso}, {Aston}, {Astone}, {Atkinson},
  {Aufmuth}, {Aulbert}, {Babak}, {Baker}, {Ballardin}, {Ballmer}, {Barker},
  {Barnum}, {Barone}, {Barr}, {Barriga}, {Barsotti}, {Barsuglia}, {Barton},
  {Bartos}, {Bassiri}, {Bastarrika}, {Bauchrowitz}, {Bauer}, {Behnke}, {Beker},
  {Benacquista}, {Bertolini}, {Betzwieser}, {Beveridge}, {Beyersdorf},
  {Bigotta}, {Bilenko}, {Billingsley}, {Birch}, {Birindelli}, {Biswas},
  {Bitossi}, {Bizouard}, {Black}, {Blackburn}, {Blackburn}, {Blair}, {Bland},
  {Blom}, {Blomberg}, {Boccara}, {Bock}, {Bodiya}, {Bondarescu}, {Bondu},
  {Bonelli}, {Bork}, {Born}, {Bose}, {Bosi}, {Boyle}, {Braccini}, {Bradaschia},
  {Brady}, {Braginsky}, {Brau}, {Breyer}, {Bridges}, {Brillet}, {Brinkmann},
  {Brisson}, {Britzger}, {Brooks}, {Brown}, {Budzy{\'n}ski}, {Bulik}, {Bulten},
  {Buonanno}, {Burguet-Castell}, {Burmeister}, {Buskulic}, {Byer}, {Cadonati},
  {Cagnoli}, {Calloni}, {Camp}, {Campagna}, {Campsie}, {Cannizzo}, {Cannon},
  {Canuel}, {Cao}, {Capano}, {Carbognani}, {Caride}, {Caudill}, {Cavagli{\`a}},
  {Cavalier}, {Cavalieri}, {Cella}, {Cepeda}, {Cesarini}, {Chalermsongsak},
  {Chalkley}, {Charlton}, {Chassande Mottin}, {Chelkowski}, {Chen},
  {Chincarini}, {Christensen}, {Chua}, {Chung}, {Clark}, {Clark}, {Clayton},
  {Cleva}, {Coccia}, {Colacino}, {Colas}, {Colla}, {Colombini}, {Conte},
  {Cook}, {Corbitt}, {Corda}, {Cornish}, {Corsi}, {Costa}, {Coulon}, {Coward},
  {Coyne}, {Creighton}, {Creighton}, {Cruise}, {Culter}, {Cumming},
  {Cunningham}, {Cuoco}, {Dahl}, {Danilishin}, {Dannenberg}, {D'Antonio},
  {Danzmann}, {Dari}, {Das}, {Dattilo}, {Daudert}, {Davier}, {Davies}, {Davis},
  {Daw}, {Day}, {Dayanga}, {De Rosa}, {DeBra}, {Degallaix}, {del Prete},
  {Dergachev}, {DeRosa}, {DeSalvo}, {Devanka}, {Dhurandhar}, {Di Fiore}, {Di
  Lieto}, {Di Palma}, {Emilio}, {Di Virgilio}, {D{\'{\i}}az}, {Dietz},
  {Donovan}, {Dooley}, {Doomes}, {Dorsher}, {Douglas}, {Drago}, {Drever},
  {Driggers}, {Dueck}, {Dumas}, {Eberle}, {Edgar}, {Edwards}, {Effler},
  {Ehrens}, {Engel}, {Etzel}, {Evans}, {Evans}, {Fafone}, {Fairhurst}, {Fan},
  {Farr}, {Fazi}, {Fehrmann}, {Feldbaum}, {Ferrante}, {Fidecaro}, {Finn},
  {Fiori}, {Flaminio}, {Flanigan}, {Flasch}, {Foley}, {Forrest}, {Forsi},
  {Fotopoulos}, {Fournier}, {Franc}, {Frasca}, {Frasconi}, {Frede}, {Frei},
  {Frei}, {Freise}, {Frey}, {Fricke}, {Friedrich}, {Fritschel}, {Frolov},
  {Fulda}, {Fyffe}, {Gammaitoni}, {Garofoli}, {Garufi}, {Gemme}, {Genin},
  {Gennai}, {Gholami}, {Ghosh}, {Giaime}, {Giampanis}, {Giardina}, {Giazotto},
  {Gill}, {Goetz}, {Goggin}, {Gonz{\'a}lez}, {Gorodetsky}, {Go{\ss}ler},
  {Gouaty}, {Graef}, {Granata}, {Grant}, {Gras}, {Gray}, {Greenhalgh},
  {Gretarsson}, {Greverie}, {Grosso}, {Grote}, {Grunewald}, {Guidi},
  {Gustafson}, {Gustafson}, {Hage}, {Hall}, {Hallam}, {Hammer}, {Hammond},
  {Hanks}, {Hanna}, {Hanson}, {Harms}, {Harry}, {Harry}, {Harstad}, {Haughian},
  {Hayama}, {Heefner}, {Heitmann}, {Hello}, {Heng}, {Heptonstall}, {Hewitson},
  {Hild}, {Hirose}, {Hoak}, {Hodge}, {Holt}, {Hosken}, {Hough}, {Howell},
  {Hoyland}, {Huet}, {Hughey}, {Husa}, {Huttner}, {Huynh-Dinh}, {Ingram},
  {Inta}, {Isogai}, {Ivanov}, {Jaranowski}, {Johnson}, {Jones}, {Jones},
  {Jones}, {Ju}, {Kalmus}, {Kalogera}, {Kandhasamy}, {Kanner}, {Katsavounidis},
  {Kawabe}, {Kawamura}, {Kawazoe}, {Kells}, {Keppel}, {Khalaidovski},
  {Khalili}, {Khazanov}, {Kim}, {Kim}, {King}, {Kinzel}, {Kissel}, {Klimenko},
  {Kondrashov}, {Kopparapu}, {Koranda}, {Kowalska}, {Kozak}, {Krause},
  {Kringel}, {Krishnamurthy}, {Krishnan}, {Kr{\'o}lak}, {Kuehn}, {Kullman},
  {Kumar}, {Kwee}, {Landry}, {Lang}, {Lantz}, {Lastzka}, {Lazzarini}, {Leaci},
  {Leong}, {Leonor}, {Leroy}, {Letendre}, {Li}, {Li}, {Lin}, {Lindquist},
  {Lockerbie}, {Lodhia}, {Lorenzini}, {Loriette}, {Lormand}, {Losurdo}, {Lu},
  {Luan}, {Lubinski}, {Lucianetti}, {L{\"u}ck}, {Lundgren}, {Machenschalk},
  {MacInnis}, {Mackowski}, {Mageswaran}, {Mailand}, {Majorana}, {Mak}, {Man},
  {Mandel}, {Mandic}, {Mantovani}, {Marchesoni}, {Marion}, {M{\'a}rka},
  {M{\'a}rka}, {Maros}, {Marque}, {Martelli}, {Martin}, {Martin}, {Marx},
  {Mason}, {Masserot}, {Matichard}, {Matone}, {Matzner}, {Mavalvala},
  {McCarthy}, {McClelland}, {McGuire}, {McIntyre}, {McIvor}, {McKechan},
  {Meadors}, {Mehmet}, {Meier}, {Melatos}, {Melissinos}, {Mendell},
  {Men{\'e}ndez}, {Mercer}, {Merill}, {Meshkov}, {Messenger}, {Meyer}, {Miao},
  {Michel}, {Milano}, {Miller}, {Minenkov}, {Mino}, {Mitra}, {Mitrofanov},
  {Mitselmakher}, {Mittleman}, {Moe}, {Mohan}, {Mohanty}, {Mohapatra},
  {Moraru}, {Moreau}, {Moreno}, {Morgado}, {Morgia}, {Morioka}, {Mors},
  {Mosca}, {Moscatelli}, {Mossavi}, {Mours}, {MowLowry}, {Mueller},
  {Mukherjee}, {Mullavey}, {M{\"u}ller-Ebhardt}, {Munch}, {Murray}, {Nash},
  {Nawrodt}, {Nelson}, {Neri}, {Newton}, {Nishizawa}, {Nocera}, {Nolting},
  {Ochsner}, {O'Dell}, {Ogin}, {Oldenburg}, {O'Reilly}, {O'Shaughnessy},
  {Osthelder}, {Ottaway}, {Ottens}, {Overmier}, {Owen}, {Page}, {Pagliaroli},
  {Palladino}, {Palomba}, {Pan}, {Pankow}, {Paoletti}, {Papa}, {Pardi},
  {Pareja}, {Parisi}, {Pasqualetti}, {Passaquieti}, {Passuello}, {Patel},
  {Pedraza}, {Pekowsky}, {Penn}, {Peralta}, {Perreca}, {Persichetti}, {Pichot},
  {Pickenpack}, {Piergiovanni}, {Pietka}, {Pinard}, {Pinto}, {Pitkin},
  {Pletsch}, {Plissi}, {Poggiani}, {Postiglione}, {Prato}, {Predoi}, {Price},
  {Prijatelj}, {Principe}, {Privitera}, {Prix}, {Prodi}, {Prokhorov},
  {Puncken}, {Punturo}, {Puppo}, {Quetschke}, {Raab}, {Rabaste}, {Rabeling},
  {Radke}, {Radkins}, {Raffai}, {Rakhmanov}, {Rankins}, {Rapagnani}, {Raymond},
  {Re}, {Reed}, {Reed}, {Regimbau}, {Reid}, {Reitze}, {Ricci}, {Riesen},
  {Riles}, {Roberts}, {Robertson}, {Robinet}, {Robinson}, {Robinson}, {Rocchi},
  {Roddy}, {R{\"o}ver}, {Rogstad}, {Rolland}, {Rollins}, {Romano}, {Romano},
  {Romie}, {Rosi{\'n}ska}, {Rowan}, {R{\"u}diger}, {Ruggi}, {Ryan}, {Sakata},
  {Sakosky}, {Salemi}, {Sammut}, {Sancho de la Jordana}, {Sandberg},
  {Sannibale}, {Santamar{\'{\i}}a}, {Santostasi}, {Saraf}, {Sassolas},
  {Sathyaprakash}, {Sato}, {Satterthwaite}, {Saulson}, {Savage}, {Schilling},
  {Schnabel}, {Schofield}, {Schulz}, {Schutz}, {Schwinberg}, {Scott}, {Scott},
  {Searle}, {Seifert}, {Sellers}, {Sengupta}, {Sentenac}, {Sergeev},
  {Shaddock}, {Shapiro}, {Shawhan}, {Shoemaker}, {Sibley}, {Siemens}, {Sigg},
  {Singer}, {Sintes}, {Skelton}, {Slagmolen}, {Slutsky}, {Smith}, {Smith},
  {Smith}, {Somiya}, {Sorazu}, {Speirits}, {Stein}, {Stein}, {Steinlechner},
  {Steplewski}, {Stochino}, {Stone}, {Strain}, {Strigin}, {Stroeer}, {Sturani},
  {Stuver}, {Summerscales}, {Sung}, {Susmithan}, {Sutton}, {Swinkels},
  {Talukder}, {Tanner}, {Tarabrin}, {Taylor}, {Taylor}, {Thomas}, {Thorne},
  {Thorne}, {Thrane}, {Th{\"u}ring}, {Titsler}, {Tokmakov}, {Toncelli},
  {Tonelli}, {Torres}, {Torrie}, {Tournefier}, {Travasso}, {Traylor}, {Trias},
  {Trummer}, {Tseng}, {Ugolini}, {Urbanek}, {Vahlbruch}, {Vaishnav}, {Vajente},
  {Vallisneri}, {van den Brand}, {Van Den Broeck}, {van der Putten}, {van der
  Sluys}, {van Veggel}, {Vass}, {Vaulin}, {Vavoulidis}, {Vecchio}, {Vedovato},
  {Veitch}, {Veitch}, {Veltkamp}, {Verkindt}, {Vetrano}, {Vicer{\'e}},
  {Villar}, {Vinet}, {Vocca}, {Vorvick}, {Vyachanin}, {Waldman}, {Wallace},
  {Wanner}, {Ward}, {Was}, {Wei}, {Weinert}, {Weinstein}, {Weiss}, {Wen},
  {Wen}, {Wessels}, {West}, {Westphal}, {Wette}, {Whelan}, {Whitcomb}, {White},
  {Whiting}, {Wilkinson}, {Willems}, {Williams}, {Willke}, {Winkelmann},
  {Winkler}, {Wipf}, {Wiseman}, {Woan}, {Wooley}, {Worden}, {Yakushin},
  {Yamamoto}, {Yamamoto}, {Yeaton-Massey}, {Yoshida}, {Yu}, {Yvert}, {Zanolin},
  {Zhang}, {Zhang}, {Zhao}, {Zotov}, {Zucker}, {Zweizig}, {The LIGO Scientific
  Collaboration}, {the Virgo Collaboration}, \& {Belczynski}}]{lig10}
{Abadie}, J., {et~al.} 2010, Classical and Quantum Gravity, 27, 173001

\bibitem[{{Belczy{\'n}ski} {et~al.}(2000){Belczy{\'n}ski}, {Bulik}, \&
  {Zbijewski}}]{bel00}
{Belczy{\'n}ski}, K., {Bulik}, T., \& {Zbijewski}, W. 2000, \aap, 355, 479

\bibitem[{{Bloom} {et~al.}(1999){Bloom}, {Sigurdsson}, \& {Pols}}]{blo99}
{Bloom}, J.~S., {Sigurdsson}, S., \& {Pols}, O.~R. 1999, \mnras, 305, 763

\bibitem[{{Blumenthal} {et~al.}(1984){Blumenthal}, {Faber}, {Primack}, \&
  {Rees}}]{blu84}
{Blumenthal}, G.~R., {Faber}, S.~M., {Primack}, J.~R., \& {Rees}, M.~J. 1984,
  \nat, 311, 517

\bibitem[{{Brandt} \& {Podsiadlowski}(1995)}]{bra95}
{Brandt}, N., \& {Podsiadlowski}, P. 1995, \mnras, 274, 461

\bibitem[{{Bulik} {et~al.}(1999){Bulik}, {Belczy{\'n}ski}, \&
  {Zbijewski}}]{bul99}
{Bulik}, T., {Belczy{\'n}ski}, K., \& {Zbijewski}, W. 1999, \mnras, 309, 629

\bibitem[{{Cutler} \& {Flanagan}(1994)}]{cut94}
{Cutler}, C., \& {Flanagan}, {\'E}.~E. 1994, \prd, 49, 2658

\bibitem[{{Fairhurst}(2009)}]{fai09}
{Fairhurst}, S. 2009, New Journal of Physics, 11, 123006

\bibitem[{{Fairhurst, S., et. al.}(2010)}]{fai10}
{Fairhurst, S., et. al.} 2010, in preparation

\bibitem[{{Fryer} \& {Kalogera}(1997)}]{fry97}
{Fryer}, C., \& {Kalogera}, V. 1997, \apj, 489, 244

\bibitem[{{Fryer} {et~al.}(1999){Fryer}, {Woosley}, \& {Hartmann}}]{fry99}
{Fryer}, C.~L., {Woosley}, S.~E., \& {Hartmann}, D.~H. 1999, \apj, 526, 152

\bibitem[{{Guetta} \& {Piran}(2005)}]{gue05}
{Guetta}, D., \& {Piran}, T. 2005, \aap, 435, 421

\bibitem[{{Hoffman} {et~al.}(2008){Hoffman}, {Martinez-Vaquero}, {Yepes}, \&
  {Gottl{\"o}ber}}]{hof08}
{Hoffman}, Y., {Martinez-Vaquero}, L.~A., {Yepes}, G., \& {Gottl{\"o}ber}, S.
  2008, \mnras, 386, 390

\bibitem[{{Hughes}(2009)}]{hug09}
{Hughes}, S.~A. 2009, \araa, 47, 107

\bibitem[{{Kulkarni}(2005)}]{kul05}
{Kulkarni}, S.~R. 2005, ArXiv Astrophysics e-prints (arXiv:astro-ph/0510256)

\bibitem[{{Lee} \& {Ramirez-Ruiz}(2007)}]{lee07}
{Lee}, W.~H., \& {Ramirez-Ruiz}, E. 2007, New Journal of Physics, 9, 17

\bibitem[{{Li} \& {Paczy{\'n}ski}(1998)}]{li98}
{Li}, L., \& {Paczy{\'n}ski}, B. 1998, \apjl, 507, L59

\bibitem[{{Madau} {et~al.}(1996){Madau}, {Ferguson}, {Dickinson}, {Giavalisco},
  {Steidel}, \& {Fruchter}}]{mad96}
{Madau}, P., {Ferguson}, H.~C., {Dickinson}, M.~E., {Giavalisco}, M.,
  {Steidel}, C.~C., \& {Fruchter}, A. 1996, \mnras, 283, 1388

\bibitem[{{Madau} {et~al.}(1998){Madau}, {Pozzetti}, \& {Dickinson}}]{mad98}
{Madau}, P., {Pozzetti}, L., \& {Dickinson}, M. 1998, \apj, 498, 106

\bibitem[{{Mandel}(2010)}]{man10}
{Mandel}, I. 2010, \prd, 81, 084029

\bibitem[{{Mandel} \& {O'Shaughnessy}(2010)}]{man10b}
{Mandel}, I., \& {O'Shaughnessy}, R. 2010, Classical and Quantum Gravity, 27,
  114007

\bibitem[{{Metzger} {et~al.}(2010){Metzger}, {Mart{\'{\i}}nez-Pinedo},
  {Darbha}, {Quataert}, {Arcones}, {Kasen}, {Thomas}, {Nugent}, {Panov}, \&
  {Zinner}}]{met10}
{Metzger}, B.~D., {et~al.} 2010, \mnras, 406, 2650

\bibitem[{{O'Shaughnessy} {et~al.}(2010){O'Shaughnessy}, {Kalogera}, \&
  {Belczynski}}]{osh10}
{O'Shaughnessy}, R., {Kalogera}, V., \& {Belczynski}, K. 2010, \apj, 716, 615

\bibitem[{{O'Shaughnessy} {et~al.}(2008){O'Shaughnessy}, {Kim}, {Kalogera}, \&
  {Belczynski}}]{osh08}
{O'Shaughnessy}, R., {Kim}, C., {Kalogera}, V., \& {Belczynski}, K. 2008, \apj,
  672, 479

\bibitem[{{Phinney}(1991)}]{phi91}
{Phinney}, E.~S. 1991, \apjl, 380, L17

\bibitem[{{Portegies Zwart} \& {Yungelson}(1998)}]{por98}
{Portegies Zwart}, S.~F., \& {Yungelson}, L.~R. 1998, \aap, 332, 173

\bibitem[{{Rosswog} \& {Ramirez-Ruiz}(2002)}]{ros02}
{Rosswog}, S., \& {Ramirez-Ruiz}, E. 2002, \mnras, 336, L7

\bibitem[{{Rosswog} {et~al.}(2003){Rosswog}, {Ramirez-Ruiz}, \&
  {Davies}}]{ros03}
{Rosswog}, S., {Ramirez-Ruiz}, E., \& {Davies}, M.~B. 2003, \mnras, 345, 1077

\bibitem[{{Spergel} {et~al.}(2007){Spergel}, {Bean}, {Dor{\'e}}, {Nolta},
  {Bennett}, {Dunkley}, {Hinshaw}, {Jarosik}, {Komatsu}, {Page}, {Peiris},
  {Verde}, {Halpern}, {Hill}, {Kogut}, {Limon}, {Meyer}, {Odegard}, {Tucker},
  {Weiland}, {Wollack}, \& {Wright}}]{spe07}
{Spergel}, D.~N., {et~al.} 2007, \apjs, 170, 377

\bibitem[{{Stadel}(2001)}]{sta01}
{Stadel}, J.~G. 2001, PhD thesis, University of Washington

\bibitem[{{van der Sluys} {et~al.}(2008){van der Sluys}, {R{\"o}ver},
  {Stroeer}, {Raymond}, {Mandel}, {Christensen}, {Kalogera}, {Meyer}, \&
  {Vecchio}}]{slu08}
{van der Sluys}, M.~V., {et~al.} 2008, \apjl, 688, L61

\bibitem[{{Zemp} {et~al.}(2009){Zemp}, {Ramirez-Ruiz}, \& {Diemand}}]{zem09}
{Zemp}, M., {Ramirez-Ruiz}, E., \& {Diemand}, J. 2009, \apjl, 705, L186

\end{thebibliography}

\clearpage


\begin{table}[h]
	\renewcommand\arraystretch{1.4}
	\centering      
	\begin{tabular}{| c | c || c | c || c | c || c | c |}
\hline
  &  & \multicolumn{2}{c||}{$M_{360}$($D_{360}$)} & \multicolumn{2}{c||}{$M_{180}$($D_{180}$)}  & \multicolumn{2}{c|}{$M_{90}$($D_{90}$)}  \\ \hline
Dist & PSF Accuracy & vs. $M_{180}$ & vs. $M_{90}$ & vs. $M_{360}$ & vs. $M_{90}$ & vs. $M_{360}$ & vs. $M_{180}$ \\
\hline \hline
\multirow{3}{*}{ $\leq$ 80Mpc}  & High &  22 & 16 & 26 & $\mbox{\scriptsize $>1000$}$ & 22 & 282 \\ \cline{2-8}    
						& Med & 73 & 39 & 35 & $\mbox{\scriptsize $>1000$}$ & 31 & 384 \\ \cline{2-8}  		
				   		& Low  & $\mbox{\scriptsize $>1000$}$ & 349 & 52 & $\mbox{\scriptsize $>1000$}$ & 50 & 881  \\
\hline \hline
\multirow{3}{*}{$\leq$ 40Mpc} & High  & 27 & 17 & 34 & $\mbox{\scriptsize $>1000$}$ & 23 & $\mbox{\scriptsize $>1000$}$  \\ \cline{2-8}
					       & Med  & 78 & 46 & 40 & $\mbox{\scriptsize $>1000$}$ & 37 & $\mbox{\scriptsize $>1000$}$ \\ \cline{2-8}
					       & Low   & 146 & 137 & 56 & $\mbox{\scriptsize $>1000$}$ & 56 & $\mbox{\scriptsize $>1000$}$  \\ \hline
					       
\end{tabular} 
	\caption{\normalfont{Number of detections required to achieve 99\% confidence in the correct model for $90\%$ ($\frac{45}{50}$) of data sets.  These results are compared between two different sample radii, and three different detector accuracies characterized by standard deviations (in distance, right ascension, declination) of:  $\sigma_{high} = \{5\%, 1^\circ, 1^\circ\}$, \hspace{0.05in} $\sigma_{med} = \{30\%, 2^\circ, 2^\circ\}$,  \hspace{0.05in}  $\sigma_{low} = \{50\%, 4^\circ, 4^\circ\}$.  Entries marked `$\mbox{\scriptsize $>1000$}$' failed to reach the desired confidence in the required number of data sets within the 1000 data points used.} }
	\label{tab_bay}
\end{table}

\end{document}